\documentclass[aps,twocolumn,showpacs,superscriptaddress]{revtex4}

\usepackage{amsmath,amssymb}
\usepackage{url}
\usepackage{graphicx,color}
\usepackage{float}

\newcommand{\ket}[1]{\left| #1 \right\rangle}
\newcommand{\bra}[1]{\left\langle #1 \right|}
\newcommand{\im}{\ensuremath{\mathrm{i}}}
\newcommand{\etal}{\mbox{\emph{et.~al.}} } 

\graphicspath{{./figures/}}

\begin{document}

 \title{Tracking spin and charge with spectroscopy in spin-polarised 1D systems}
 \author{Tobias Ulbricht}
 \affiliation{                    
   Institut f\"ur Theorie der Kondensierten Materie
   - Karlsruher Institut f\"ur Technologie, 76131 Karlsruhe, Germany
 }
 
 \author{Peter Schmitteckert}
 \affiliation{                    
   Institut f\"ur Nanotechnologie - Karlsruher Institut f\"ur Technologie, 76131 Karlsruhe, Germany
 }
 
 \begin{abstract}
   We calculate the spectral function of a one-dimensional strongly
   interacting chain of fermions, where the response can be well
   understood in terms of spinon and holon excitations. %
   Upon increasing the spin imbalance between the spin species, we
   observe the single-electron response of the fully polarised system
   to emanate from the holon peak while the spinon response
   vanishes. %
   For experimental setups that probe one-dimensional properties, we
   propose this method as an additional generic tool to aid the
   identification of spectral structures, e.g.~in ARPES
   measurements. %
   We show that this applies even to trapped systems having cold
   atomic gas experiments in mind. %
 \end{abstract}
 \pacs{71.10.Pm,71.10.Fd,72.15.Nj}

\maketitle

\section{Introduction}
The interest in one-dimensional structures is unwaning. %
For the most part, one wants to understand the special nature of
one-dimensional materials and verify theoretical models thereof. %
Sometimes, it is believed to help solving puzzles of two-dimensional
systems
\cite{Graf_Gweon_McElroy_Zhou_Jozwiak_Rotenberg_Bill_Sasagawa_Eisaki_Uchida_Takagi_Lee_Lanzara_PRL2007}. %
The Luttinger liquid \cite{Haldane_JPC1981}, as the accepted effective
low-energy theory, suggests that the elementary excitations are
bosonic collective excitations, in contrast to the quasi-particles of
Fermi liquid theory in higher dimensions. %
The fundamental collective excitations either carry charge $\pm e$ and
no spin (holons/antiholons) or carry the spin $\frac{1}{2}$ and no
charge (spinons). %
This so-called spin-charge separation becomes explicit in the exact
solution of the Hubbard model
\cite{Essler_Frahm_Gohmann_Klumper_Korepin_BOOK2005} by Lieb and Wu
\cite{Lieb_Wu_PRL1968} using the Bethe ansatz. %

Therefore, physicists look for experimental evidence of spin-charge
separation in quasi-one-dimensional materials
\cite{Halperin_JAP2007}. %
Recent promising experiments include the momentum-conserved tunnelling
measurements between parallel GaAs/AlGaAs wires
\cite{Auslaender_Steinberg_Yacoby_Tserkovnyak_Halperin_Baldwin_Pfeiffer_West_S2005}
and conductance measurements on single wall carbon nanotubes
\cite{Bockrath_Cobden_Lu_Rinzler_Smalley_Balents_McEuen_N1999}. %
Further, theoretical approaches propose to use magneto-tunneling
\cite{Altland_Barnes_Hekking_Schofield_PRL1999} or a transport setup
\cite{Ulbricht_Schmitteckert_EPL2009} on quantum wires, or to extract
the information from ultra-cold atomic gases on a lattice
\cite{Kleine_Kollath_McCulloch_Giamarchi_Schollwock_PRA2008}. %
However, experimental techniques of angle resolved photoelectron
spectroscopy (ARPES) have vastly improved over the last years,
providing momentum-resolved spectral densities of occupied states in
the valence bands. %
The most prominent experiments using ARPES are those of Kim \etal
\cite{Kim_Koh_Rotenberg_Oh_Eisaki_Motoyama_Uchida_Tohyama_Maekawa_Shen_Kim_NPhys2006}
on SrCuO$_{2}$ and recent measurements by Claessen \etal
\cite{Claessen_Sing_Schwingenschlogl_Blaha_Dressel_Jacobsen_PRL2002}
on the organic complex TTF-TCNQ, but also one-dimensional metal atom
chains, e.g.~Au on a Si(111) substrate
\cite{Segovia_Purdie_Hengsberger_Baer_N1999} or Au on Ge(001)
\cite{Schafer_Blumenstein_Meyer_Wisniewski_Claessen_PRL2008} are under
investigation. %
Every so often, the interpretation of experiments are critically
discussed and challenged to be ambiguous, especially when spin and
charge scales can not be determined independently (see, e.g.
\cite{Kim_Koh_Rotenberg_Oh_Eisaki_Motoyama_Uchida_Tohyama_Maekawa_Shen_Kim_NPhys2006}). %

ARPES experiments deliver momentum-resolved spectral densities
$\mathcal{A}(k,\omega) = \sum_{\sigma}\mathcal{A}^{<}_{\sigma}$, which
probe the response of the system upon removal of an electron, with
\begin{align}
  \label{eq:specf}
  \pi \mathcal{A}^{<}_{\sigma} (k,\omega) &\propto \text{Im} 
  \left\{ \bra{\Psi_{0}} c_{k\sigma}^{\dagger} \frac{1}{(E_{0} - H -\omega - \im \eta)} c_{k\sigma}^{} \ket{\Psi_{0}} \right\},
\end{align}
being the imaginary part of the one-particle Greens function, where
$H$ denotes the Hamiltonian of the system, $E_0$ its ground state
energy corresponding to the ground state $\ket{\Psi_{0}}$,
$c^{(\dagger)}_{k\sigma}$ are the free single particle operators
corresponding to a Bloch state with momentum $k$ and spin $\sigma$ and
$\eta$ is the broadening in finite systems. %
The spectral function encodes valuable information about the
elementary excitations, e.g. as a function of $\omega$ at $k=0$, it
reveals the energetically separated spinon and holon excitation
response. %

An additional homogeneous magnetic field breaks spin invariance and
favours the population of one of the spin species, driving the spin
degrees of freedom into alignment. %
In this work, we propose that tuning the spin polarisation by such a
magnetic field can be used as a tool to trace the spinon-like and
holon-like excitations that are found (or claimed to be found) at zero
field. %
We will show that the trajectories of the spectral features allow a
unique determination of the holon and thus allows for a definite
identification of spinon and holon at zero magnetic field. %
Especially, this argument relieves experiments of the pressure to
measure spin and charge energy scales independently. %
Using a density matrix renormalisation group (DMRG) method we are the
first to calculate the spectral function \eqref{eq:specf} in a
magnetic field without restriction on energy or momentum. %
This is shown for the Mott-insulating Hubbard model, the Hubbard model
in the metallic phase, and the Hubbard model in a trapping potential,
comparable to cold atom gas experiments. %
As a by-product, we find uncommon signatures that shed light on the
unexplored nature of the elementary excitations in the Hubbard model
at finite magnetic. %

\section{Zero magnetic field}
Starting point of our investigation is the zero-field spectral
function of one-dimensional interacting electrons. %
Even for integrable models, exact analytical treatments of the
spectral function are limited. %
For a spinful Luttinger liquid model, where the spin and charge
sectors decouple completely, Meden and Sch\"onhammer
\cite{Meden_Schonhammer_PRB1992} and Voit \cite{Voit_JPC1993} derived
algebraic singularities in $\mathcal{A}(k,\omega)$ corresponding to
the holon and spinon response, which are, due to the nature of the
Luttinger liquid limited to a region around $\omega=0$. %
For the simplest non-trivial, but experimentally relevant microscopic
model, the Hubbard model, the celebrated Bethe ansatz offers an exact
solution by Lieb and Wu \cite{Lieb_Wu_PRL1968} and yields the
collective mode's excitation spectrum for all values of interaction
and filling. %
Unfortunately, it is very tedious to extract the spectral function
from the Lieb-Wu equations, but the spinon and holon dispersion leave
a distinct footprint in the spectral function of an electron
scattering state
\cite{Essler_Frahm_Gohmann_Klumper_Korepin_BOOK2005}. %
Only recent developments using a pseudofermion dynamical theory (see
Ref.~\cite{Carmelo_Penc_EPJB2006} and references therein) facilitated
the calculation of the spectral function in this model
\cite{Bozi_Carmelo_Penc_Sacramento_JPC2008} without the limitation of
being perturbative in large or small interaction or the restriction to
a low-energy effective theory. %
Further, in certain Mott-insulator or charge-density wave insulator
systems, the spectral function was extracted adopting a field
theoretical description in the Luttinger framework
\cite{Essler_Tsvelik_PRL2002, Essler_Tsvelik_PRL2003,
  Schuricht_Essler_Jaefari_Fradkin_PRL2008}. %

Numerical approaches are numerous, too. %
Exact diagonalisation, as always restricted to small system sizes, was
used e.g.~on the $t$-$J$ model by Eder and Ohta
\cite{Eder_Ohta_PRB1997} to compare 2D and 1D spectral features. %
Quantum monte carlo methods, generically suffering from the sign
problem, succeeded \cite{Senechal_Perez_Pioro-Ladriere_PRL2000,
  Zacher_Arrigoni_Hanke_Schrieffer_PRB1998} in resolving the main
spectral features of the Hubbard model as predicted by the Bethe
ansatz estimates. %
One advantage over the DMRG method is the easy extension to two
dimensions also studied by S\'en\'echal \etal
\cite{Senechal_Perez_Pioro-Ladriere_PRL2000}. %
Finally, the DMRG method \cite{White_PRL1992} (see
Ref.~\cite{Noack_Manmana_AIPCP2005}
for reviews) and its extensions allow for an error-controlled and
efficient calculation of dynamical quantities
\cite{Jeckelmann_PTPS2008}. %
Especially Jeckelmann and co-workers did extensive and accurate
studies of the spectral properties of half-filled Hubbard model
\cite{Jeckelmann_Benthien_Book2008}, the Hubbard model off
half-filling \cite{Benthien_Gebhard_Jeckelmann_PRL2004}, and the
extended Hubbard model (with nearest neighbour interaction)
\cite{Benthien_Jeckelmann_PRB2007}. %
The spectral features of TTF-TCNQ
\cite{Claessen_Sing_Schwingenschlogl_Blaha_Dressel_Jacobsen_PRL2002},
e.g., seem to fit on the simple one band Hubbard model, which was
numerically established by Benthien \etal
\cite{Benthien_Gebhard_Jeckelmann_PRL2004} and lately also by Bozi
\etal \cite{Bozi_Carmelo_Penc_Sacramento_JPC2008} using the
pseudofermion method, but fail completely with a band description
using density functional theory
\cite{Claessen_Sing_Schwingenschlogl_Blaha_Dressel_Jacobsen_PRL2002}. %
Relevant to the upcoming discussion is that around the $\Gamma$-point
the spectral response $\mathcal{A}(\omega)$ displays (beside an
incoherent continuum of spinon-holon excitations) a two-peak structure
corresponding to holons at lower and spinons at higher energies. %

\section{Finite magnetic field}
The Hubbard model in a magnetic field was analysed as early as 1990 by
Frahm and Korepin \cite{Frahm_Korepin_PRB1990, Frahm_Korepin_PRB1991}
with the focus on the asymptotics of the Greens functions in time and
space. %
Even if we know of no exhaustive treatment for $\mathcal{A}(k,\omega)$
for the polarised Hubbard model or other integrable models (see
\cite{Frahm_Vekua_JSMTE2008}),
at least the low-energy behavior for all momenta (corresponding to a
momentum distribution curve in ARPES) was considered using a
pseudofermion formulation \cite{Carmelo_Guinea_Sacramento_PRL1997} and
Ref. \cite{Frahm_Korepin_PRB1991} estimated the behavior in the large
$U$ limit for special values of $\omega$ and $k$. %
Although a recent low-energy effective field theory by Zhao and Liu
\cite{Zhao_Liu_PRA2008} with attractive interaction and spin imbalance
aims at a different direction (Fulde-Ferrell-Larkin-Ovchinnikov state
vs. Fermi liquid), they note that the polarisation introduces terms
that couple the spin and the charge sector. %
Finally, the idea of detecting spin-charge separation by splitting the
spectral response with a magnetic field was also formulated within
Luttinger liquid theory by Rabello and Si \cite{Rabello_Si_EPL2002}. %
Being restricted to low-lying excitations, their proposal requires an
extreme resolution to resolve the splitting of spectral features. %

In contrast, we can calculate the spectral response
$\mathcal{A}(k,\omega)$ at any energy, momentum and interaction
strength. %
As proposed, adding an external homogeneous magnetic field will
polarise the system. %
At zero temperature there will be a critical magnetic field
$B_{\text{cr}}$, above which the minority spin species will not be
occupied. %
Assuming this regime is accessible in experiment, instead of the two
peaks, one will recover the quasi-particle ($\delta$-like) response,
since the majority spin type now represents a non-interacting electron
gas. %
Thus, we can freely tune between a free and an interacting electron
gas response. %
Going from full to zero polarisation by turning down the magnetic
field, we do not a priori know how the spectral response of an
electron relates to the spectral responses of the charge and spin
sector. %
Is there a smooth transition? %
Does the electronic peak split up into charge and spin? %
Or does it vanish, while the spinon and holon features turn up
completely unrelated? %

\section{Spin polarisation in the Hubbard model}
To answer this question, we calculate the spectral function
\eqref{eq:specf} for the Hubbard model
\begin{equation*}
  \label{eq:hubbard}
  H =
  -t\sum_{x,\sigma=\uparrow,\downarrow} \left( c^{\dagger}_{x,\sigma} c^{}_{x-1,\sigma} + \text{h.c.} \right) +  U \sum_x
   n_{x,\uparrow} n_{x,\downarrow} ,
\end{equation*}
where $U=4t$ is the Hubbard interaction strength,
$c^{(\dagger)}_{x,\sigma}$ are the local single particle fermion
operators and $n_{x,\sigma}$ is the local density of spin type
$\sigma$. %
We use a lattice with hard-wall boundary conditions on $M$ sites with
$N = N_{\uparrow} + N_{\downarrow}$ electrons. %
For the evaluation of the resolvent we employ a preconditioned
Krylov-based correction vector method \cite{Kuhner_White_PRB1999,
  Ramasesha_JCC1990}. %
We use the single particle operators corresponding to the
particle-in-a-box solutions $c_{k,\sigma} = \sqrt{\frac{2}{M+1}}
\sum_{x} \sin (k x) c_{x,\sigma}$, with $k=\pi l /(M+1), l \in
\{1,\dots M\}$, which are known to work well for finite systems
\cite{Jeckelmann_PTPS2008} and are equivalent to the Bloch state
operators in the thermodynamic limit. %
We will calculate the energy distribution curve $\mathcal{A}(\omega)$
for the smallest value possible in this expansion $k=\pi/(M+1)$, since
there the energy scales of spinon and holon are maximally spearated. %
Energy is measured in units of the hopping parameter $t$. %
The artificial broadening $\eta$ in \eqref{eq:specf} is chosen to be
$0.1$, which is larger than the level spacing but small enough to
resolve holon and spinon features. %
The resolution in the energy is $\Delta \omega = 0.05$. %
One can attain the limit $\eta \rightarrow +0$ either by a numerically
unstable deconvolution \cite{Raas_THESIS2005}
or by treating $\eta$ as a part of the self-energy and subtracting it
from the imaginary part of the inverse of the Greens function
\cite{Schmitteckert_UNPUB2009}. %
Here we use the latter. %
For the figures, a b-splined curve was used to resample to a higher
resolution in the frequency. %
The DMRG cut-off is determined by a desired constant discarded entropy
during truncation of $0.001$ and up to $1400$ density eigenstates are
used. %
\begin{figure}[ht]
  \centering
    \includegraphics[width=1.0\linewidth]{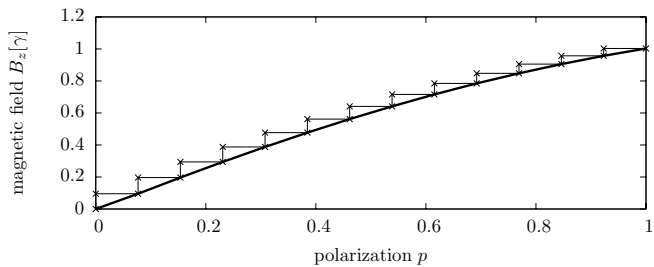}
    \caption{Dependence of $B_{z}$ in units of the gyromagnetic ratio
      $\gamma$ on the polarisation in the $U=4t$ Hubbard model for
      density $n=26/32$. %
    }
  \label{fig:scs_mag_pol}
\end{figure}

The spin polarisation is given by $p =
\frac{|N_{\uparrow}-N_{\downarrow}|}{N_{\uparrow}+N_{\downarrow}}$,
and instead of adding an external magnetic field $B$ to
\eqref{eq:specf}, we calculate in a canonical ensemble and change the
number of up ($N_{\uparrow}$) and down electrons $N_{\downarrow}$. %
From the ground state energy for each polarisation we can derive the
magnetic field energy contribution $\gamma S_{z} B_z =
E_{\text{GS}}(p) - E_{\text{GS}}(p=0)$. %
Using $S_z = \frac{1}{2} p N$, we show the dependence of $B_{z}$ on
the polarisation in Fig.~\ref{fig:scs_mag_pol} as a step function
indicating the range of possible magnetic field strengths for each
polarisation. %

Examining the metallic phase of the Hubbard model for $M=32$ sites, we
keep the density fixed at $n=n_{\uparrow} + n_{\downarrow} = 26/32 =
0.8125$, while the polarisation is stepwise tuned from 0 to 1 by
increasing $N_{\uparrow}$ and decreasing $N_{\downarrow}$.
\begin{figure}[ht]
  \centering
    \includegraphics[width=1.0\linewidth]{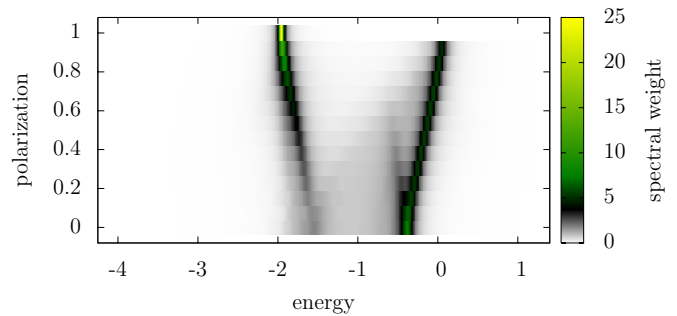}
    \caption{ %
      (Colour online) Single particle spectral function $\pi
      \mathcal{A}^{<}(k\approx 0,\omega)$ of the $U=4t$ Hubbard model
      for different polarisations as a colour-coded intensity plot,
      here for the metallic system with average density $n=26/32$. %
    }
  \label{fig:scs_sf_pol}
\end{figure}
Fig. \ref{fig:scs_sf_pol} shows an intensity plot of $\pi
\mathcal{A}(k\approx 0,\omega)$, where the $\eta$-broadened Lorentz
peak was deconvoluted in a way discussed above, retaining a residual
width of $0.02$ for visibility. %
\section{Results}
\begin{figure}[ht]
  \centering
    \includegraphics[width=1.0\linewidth]{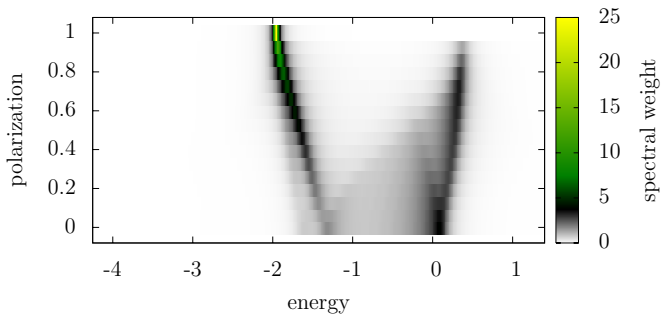}
    \caption{ %
      (Colour online) Single particle spectral function as in
      Fig.~\ref{fig:scs_sf_pol} but for the Mott-insulating state with
      density $n=1$. %
    }
  \label{fig:scs_sf_pol_mott}
\end{figure}

On the one hand, for full polarisation we recover the $\delta$-like
single particle response centered at the excitation energy
$\varepsilon(k\approx 0) = -2t$. %
Upon turning down the magnetic field, the single particle peak
broadens while decreasing in height and shifting in energy. %
For zero polarisation, we recover the established holon peak. %
Thus, the trajectory for the holon-like excitation is smoothly
transforming from the single electron peak into the holon peak. %
On the other hand, the spinon peak at $p=0$ has a more interesting
behavior upon increasing the polarisation. %
It has a higher weight than the holon peak at $p=0$, but it shifts in
energy and loses weight in accordance with the holon-like peak gaining
weight. %
While the shoulder left of the holon-like trajectory can be attributed
to the shadow band response due to $k$ not strictly zero, there is an
unexpected bifurcation in the spinon-like trajectory giving rise to a
second spinon-like excitation at lower energies and finite field. %
Its weight vanishes at some finite polarisation while the main
excitation's weight vanishes only above the critical field. %
Also, the spinon-holon continuum at intermediate energies develops a
shoulder away from the holon-like trajectory, vanishing together with
and at the spinon-like trajectory. %
Note that we find the same general behavior for the half-filled
Hubbard model, a Mott-insulator (Fig.~\ref{fig:scs_sf_pol_mott}). %
We checked for a particular polarisation for $M=32,64,96$ sized
systems that these general features are not finite size effects. %

\section{Hubbard model in a trap}
We add a harmonic trapping potential with $V_{\text{pot}} > 0$ centered
at $x_{0}=M/2$ to the Hubbard model, having then
\begin{equation*}
  \label{eq:hubbard_trap}
  H_{\text{trap}} = H + V_{\text{pot}}\sum_{i,\sigma}n_{i,\sigma} (x_{i} - x_{0})^{2}.
\end{equation*}
The motivation is rooted in the anticipated modelling of a Hubbard
lattice model with ultra-cold fermion gases, which might involve such
a trap. %
The single particle wave functions of the non-interacting problem
($U=0$) for the energetically lowest states are well described by
Hermite functions, if the trap is not too deep ($V_{\text{pot}}
(M/2)^{2} \sim t$). %
This leads us to use the operators
\begin{equation*}
  \label{eq:scs_trap_solution}
  c_{m,\sigma} = \pi^{-\frac{1}{4}}\sqrt{\frac{g}{2^{m}m!}}
  \sum_{x} H_{m}(g(x-x_{0})) \text{e}^{-\frac{1}{2} g^{2}(x-x_{0})^{2}} c_{x,\sigma}
\end{equation*}
as single particle operators in the spectral function
\eqref{eq:specf}, where $g$ is a potential dependent normalisation
constant and $H_{m}(x)$ is the $m$th Hermite polynomial. %
The level $m$ now takes the role of the momentum $k$ in
$\mathcal{A}(k,\omega)$. %
We used a system of $M=50$ sites with $N=22$, $V_{\text{pot}}=0.01$
and tested for the removal of the lowest single particle eigenstate
$m=0$ with otherwise identical parameters. %
\begin{figure}[t]
  \centering
  \includegraphics[width=1.0\linewidth]{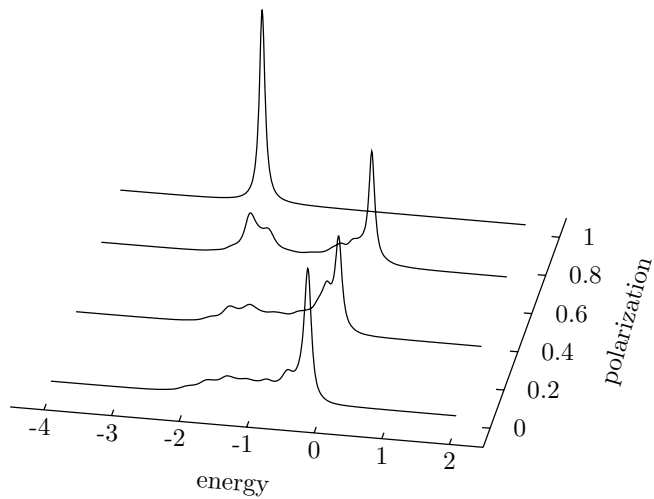}
  \caption{ Single particle spectral function in the $U=4$ Hubbard
    model in a trap for strength $V_{\text{pot}}=0.01$ for density
    $n=11/50$ for different polarisations as a profile plot. %
  }
  \label{fig:scs_sf_pol_trap}
\end{figure}
In Fig.~\ref{fig:scs_sf_pol_trap} we have plotted the spectral
function $\pi \mathcal{A}(0,\omega)$ for selected polarisations. %
Even if the terminology ``spinon'' and ``holons'' does not apply in
the trapped system, we recover the same overall picture of emerging
features as in the plain Hubbard model. %
The bifurcation exists as well, but can be seen more clearly when
looking at the polarised response $\mathcal{A}^{<}_{\uparrow}$ (not
shown). %
Finally, preliminary calculations on an extended Hubbard model with
parameters as in \cite{Benthien_Jeckelmann_PRB2007} show again a
similar overall picture. %

\section{Conclusion and Application}
The main message of our result is that a magnetic field can be used to
uniquely distinguish the holon and the spinon peak of the Hubbard
model and thus identify the spin-charge separation. %
Measuring spin and charge energy scales independently is not necessary
in this szenario. %
From zero to full spin polarisation, the spectral features of the
charge sector continuously transforms into the electronic excitation
spectrum, while spectral features of the spin sectors stay separate
and vanish. %
There are, however, more facets. %
Firstly, our results for the trapped Hubbard model can encourage
experimentalist to realise the Hubbard model on a lattice with
ultra-cold gases, since even with a trap, the basic notions of spinon
and holon physics remain. %
Secondly, in view of the results for the trapped Hubbard model and the
extended Hubbard model, we conjecture that this behavior is generic
for a spinful Luttinger Liquid. %
For integrable models this may be even analytically accessible, at
least by looking at the excitation spectrum for finite magnetic
field. %
Surprisingly, we found a bifurcation in the spinon-like response at a
finite magnetic field, suggesting an underlying elementary excitation
structure that is different from a simple spinon-holon picture. %
Further numerical results will be published elsewhere while, again,
Bethe ansatz or other analytical tools may be able to reveal the
nature of our finding. %
Finally, in view of experiments on 2D high-temperature superconductor
materials
\cite{Graf_Gweon_McElroy_Zhou_Jozwiak_Rotenberg_Bill_Sasagawa_Eisaki_Uchida_Takagi_Lee_Lanzara_PRL2007}
trying to extract 1D interacting properties and in view of the rising
interest in the FFLO state \cite{Moreo_Scalapino_PRL2007,
  Zhao_Liu_PRA2008} for attractive interaction and imbalanced systems,
we find a broad range of applications and areas that may benefit from
our proposal. %

\begin{acknowledgments}
  We thank \mbox{Sam} \mbox{Carr}, \mbox{Holger} \mbox{Schmidt},
  \mbox{Dirk} \mbox{Schuricht}, \mbox{Ronny} \mbox{Thomale} and
  \mbox{Peter} \mbox{W\"olfle} for valuable discussions and we
  acknowledge the support by the Center for Functional Nanostructures
  (CFN), project B2.10. %
\end{acknowledgments}

\end{document}